\documentstyle[epsfig,11pt]{article}
\def\f{\frac}

\def\p{\partial}
\def\l{\left}
\def\r{\right}
\def\be{\begin{equation}}
\def\ee{\end{equation}}

\def\ni{\noindent}

\def\beb{\begin{thebibliography}{99}}
\def\eeb{\end{thebibliography}}

\def\non{\nonumber}

\def\barr{\begin{array}}
\def\earr{\end{array}}
\def\lang{\langle}
\def\rang{\rangle}
\def\bearr{\begin{eqnarray}}
\def\eearr{\end{eqnarray}}
\def\bearrs{\begin{eqnarray*}}
\def\eearrs{\end{eqnarray*}}
\begin{document}
\ni
\centerline{\bf { Pre-Equilibrium Evolution Of QCD Plasma}} \\
\centerline{\bf { An Appraisal}} \\
\ni
\centerline{\sl Avijit K. Ganguly\footnote{avijit@cts.iisc.ernet.in}}\\
\centerline{\small Center for Theoretical Studies, Indian Institute of Science,
Bangalore}\\ 
\baselineskip=11pt
\section{Introduction}
\label{s_intro}
\noindent
The study of strongly interacting matter at high temperature and  density 
had attracted the interest of people all around the globe over the past couple
of years.Now with the availability of the new experimental facilities such as
the ep collider of HERA at DESY, Fermi-Lab Tevatron ( $ p\overline p$, p~A 
collision ),
BNL Relativistic Heavy Ion Collider( RHIC)  and LHC of CERN; there is a renewed 
interest in the field--from experimental as well as theoretical point of view to 
observe as well as theoreticaly model  the production of Quark Gluon Plasma 
--the elusive state of strongly interacting matter-- that is supposed to exist
in a chiral symmetry restored and deconfined phase at sufficiently high 
temperature or density or both together.So it is expected that with the
 availability of the new 
experimental data, it will be possible to pin down the uncertainities that 
existed in the previous estimates and hence provide a better model of the 
physics involved.The interest in these studies stem from both physics as well 
as astro-physics point of view, e.g to understand the structure of the 
interior of a neutron star, millisecond pulsar or for that matter in the early 
universe cosmology, when the primordial plasma was believed to exist in the form 
of quark gluon soup all around the universe. A study  of this phase of matter in
laboratory conditions is expected to shed some light on these issues and hence
to the understanding of the universe around us. But to observe or detect the QCD
plasma produced in the laboratory one needs to have a proper estimate of the energy 
density per unit voloume, initial temperature , entropy density and the number 
density of the produced partons \cite{satz}. More over the further space time 
evolution of the plasma as well as the experimental signatures also depends
on these quantities. Therefore it is crucial to have a
proper understanding of the Pre-equilibrium production phase of the QCD plasma. 
%\ni
The essential goal of these studies is to understand how in a nucleus nucleus
collision, the coherent distribution of partons in each nucleus
evolve to a highly incoherent state, 
and finally through self interaction generates a collective QCD excitation 
--that with  the passage of time evolves and re-hadronizes before getting
captured in the detectors. Generally, the pre-equilibrium  phase is supposed to
start from the time of the nuclear contact and last till the  system 
has thermalised. 
As has already been mentioned,all the planned experiments are expected to cover 
different energy scales  ranging from 200 GeV to 100Tev, so as one moves from
 the SPS to LHC energy scale  different paradigms of QCD opens up,  from 
non-perturbative  to perturbative!
Since the time taken by the nuclei to pass through each other in LHC scale 
$\sim $ 0.0005 fm/c and the time of thermalisation( for gluons) ia around
0.25 fm/c, the system spends an appreciable amount of time in the pre-equilibrium
phase. Therefore, we in our "qualitative"
description of the  pre-equilibrium phase of QGP---will start from the  
low energy scale where the non-perturbative QCD mechanisms are dominant  and 
move towards the high energy scale where the pQCD based processes  dominates the 
scenario.
%%%%%%%%%%%%%%%%%%%%%%%%%%%%%%%%%%%%%%%%%%%%%%%%%%%%%%%%%%%%%%%%%%%%%%%%%%%%
\subsection{Non-Perturbative Domain: The Flux Tube Model}
%%%%%%%%%%%%%%%%%%%%%%%%%%%%%%%%%%%%%%%%%%%%%%%%%%%%%%%%%%%%%%%%%%%%%%%%%%%%
\noindent
The idea of creating QGP  throgh flux tube decay  owes its existence to
Low and Nussinov \cite{Low}. According to this model in a relativistic heavy ion 
collision, as the target and the projectile nucleui passes through each other,
due to multiple exchange of soft gluons 
they get color charged--creating  a strong chromo-electric field between them.
This field pulls out  $q$, ${\bar{q}}$ pairs from the vacuum by tunneling.
A simple kinematic consideration shows, to create a pair of particles 
from vacuum the amount of energy required is$ \sim 2m $ on the other hand as 
these charged particles move apart over a  characteristic length scale, say 
Compton  wave length, the energy gained by the system is $ \sim 2 E/m $. Now 
one can deduce from a simple energy conservation argument that critical 
strength field strength required to produce pairs from vacuum is $ E_c\sim~~2m^2
\cite{ganguly1}$.In technical language it is said--if the strength of this
 electric field $gE$ exceeds a critical value $ gE_c=2m^2 
$, the vacuum becomes unstable against production  of $q $  ${\bar{q}}$
pairs of mass m each.
Since the system looks like a parallel plate capacitor ,this model is also known as
capacitor plate model.In this model the electric field is
a c number field and assumed to be  
constant all through out. The rate of pair production per 
unit volume and per unit time is,
\be 
 W_p = \frac {\alpha E^2}{\pi^2} e^{\frac{- \pi m^2}{gE}}
\label{productionrate}
\ee
\noindent
On the other hand because of self interacting nature of the non-Abelian gauge
fields--the color-electric field set up between the receding nuclei can
oscillate in time. In fact in this model,  the surface charge density on the nuclear plates, in the transverse 
directions are imagined to be homogeneous, with plates extended up to infinity.
As the plates move  apart, one can find out the field
configuration in the central Baryon free region, by solving the source-less
Yang-Mills (Y.M) equations.  In an idealised situation one can study
the system in ($1+1$) dimension with an axial gauge choice to get,
\begin{eqnarray}
 A_0^1
&=&\l[{\beta_1}+{\beta_1}{\beta_3}\l[\ e^{i\omega t}+e^{-i\omega t}\r]
-{i \omega}{\beta_2}\l[\ e^{i\omega t}-e^{-i\omega t}\r] \r]z+{\beta_1} 
\nonumber \\
 A_0^2
&=&\l[{\beta_2}+{\beta_2}{\beta_3}\l[\ e^{i\omega t}+e^{-i\omega t}\r]
+{i \omega}{\beta_1}\l[\ e^{i\omega t}-e^{-i\omega t}\r] \r]z+{\beta_2}
\nonumber \\
 A_0^3
&=&\l[{\beta_3}+{{\beta_3}^2}\l[\ e^{i\omega t}+e^{-i\omega t}\r]
-{{\omega}^2}\l[\ e^{i\omega t}+e^{-i\omega t}\r] \r]z+{\beta_3} 
.\end{eqnarray}
Here since $\omega = \sqrt{ {\beta{_1}}^2+{\beta{_2}}^2+{\beta{_3}}^2}$, the 
frequency of oscillation is seen to be dependent on amplitude. 
This is the  motivation to re-investigate Schwinger mechanism
in an oscillating field. 
Since finding out $W_p$ using these exact solution, is a formidable task
an analysis was carried out in \cite{ganguly2}, to see effect of a sinusoidally 
time varying field on pair production.The result showed that, the production
probability--due to  the presence of the oscillating field goes like
\be 
W_g \sim \frac{\alpha_s E^2}{8}\left[ \frac{(gE)^2}{4 m^2 \omega_0^2} 
\right]^{\frac{2m}{\omega_0}}
\ee
Since $\omega_0$ is the energy associated with each quanta of the oscillating
gluonic field, so m divided by ${\omega_0}$ is the number of gluon quanta
required to produce a pair, hence it is called multi-gluon  ionization process. 
This result is valid even  when $ E \ll E_{c}$ and dominates over that of 
the tunneling rate. A comparison  with the  perturbative approach
also established \cite{ganguly2} that for constituent quark  mass 
the multi Gluon ionisation process is dominant over the other two.
The effect of time variation of the electric field  
was also recently considered in \cite{bhalerao}, for an exponentially  
decaying Electric filed( Decay due to depletion of the electric 
field caused by  particle production)and the conclusion was, time dependent 
field increases the production rate.
 We will discuss   this  in an appropriate section. Before closing the
discussion on nonperturbative production mechanisms its worth mentioning
that using similar techniques,  one can estimate the decay rate for Gluons
too and for an constant external electric field,the rate was found to be 
\cite{Gyu} similar to the quark case with mass set to equal to zero and a
 different color factor sitting in front. 
Few other interesting extensions of this model have been performed 
in\cite{finitesize}  where the effects of finite size correction and the effect 
of a confinning potential have been studied separately.
\section{Hard and semi Hard processes}
%\subsection{Parton Cascade Model}
Having outlined the non-perturbative production mechanisms,
we will discuss here the perturbative mechanisms, responsible for
the production of quarks and gluons at a higher energy scale. 
The idea here  is, right after collision the coherent distribution 
of the partons over the nuclei gets disrupted, as the valance partons of each 
individual nuclei starts  colliding with that  of the other--producing an 
incoherent bunch of partonic shower that--decays  either forming space like 
 or time like cascades, producing a mini-jetty enviornment . 
 The important  input in this theory comes from 
the nucleonic  structure functions measured or predicted for different values
of the fractional longitudinal momentum that the colliding parton in question 
is carrying, as known in the literature as Bjorken  x and the energy scale of 
collision $ Q^2$. The quantity of interest in this model is the jet cross 
section defined in terms of the  double differential cross section for parton 
parton collisions\cite{eskola}, as,
\be
\frac{d~ \sigma_{jet}}{d~ p_{T}^2 d~y_{1}d~y_{2}} = K \Sigma_{a,b}
 x_{1}f_{a} (x_{1}, p_{T}^2) x_{2}f_{b} (x_{2}, p_{T}^2)
\frac{d~ \sigma_{jet}}{dt}
\label{Jet_crossection}
\ee
Here the subscripts a and b refer to  the different species of partons. The 
Bjorken variable $x_{1}$ and $x_{2}$ denotes the longitudinal momentum fraction
carried by the partons of respective nucleon with $ f_{a}$  and $f_{b}$ 
as their structure functions. Also  $\frac{d~ \sigma_{jet}}{dt}$ is the  
double-differential cross section associated with partons of type a and b.
The energy density is estimated  from eqn.(\ref{Jet_crossection}) by using the
relation;
 \be
\epsilon_{h} = \frac{d~E_{T}^{AA}}{d~ Y \pi R_{A}^2 \tau_{h}} =
 \frac {T_{AA}(b)}{\pi R_{A}^2 \tau_{h} }
\sigma_{jet} \langle E_{T}^{pp} \rangle
\ee
With $ T_{AA}(b) $ as the impact parameter and
$ \sigma_{jet} \langle E_{T}^{pp} \rangle $  energy moment of the cross section.
The parton parton subprocess cross 
section  used here has a pathological divergence, that goes as 
 $ \sim {p_{t}}^{-4} $. In order to get rid of this situation 
a "saturation criteria"  \cite{saturation} is usually invoked, that gives
$p_{0} \simeq 2GeV$ for the LHC and $p_{0} \simeq 1GeV $
 for RHIC for Pb on Pb collision at zero impact 
parameter. This model predicts that the number density of gluons will 
out number the  quarks.The  mean transverse energy of gluons found to be 
$\sim $3 GeV  with the  initial temperature $\sim$ 1.1 GeV.
% We will discuss these things in the proper perspective in 
%the discussion section.   
\subsection{McLerran Venugopalan Model}
\noindent
Other than these two  complementary processes of QGP production at
RHIC and LHC energy scales, there also exists a third scenario that 
recently was pointed out by  L.McLarren  and Raju VenuGopalan in a series of 
papers \cite{larry}, where it was argued that there is also another
source of mid rapidity gluons in the central rapidity region of large "A"
 nucleus nucleus collision.
\noindent
Their argument was, as  the surface density of
 mean charge squared fluctuation i.e $\mu^2$ goes as,
 $ \mu^2 =1.1 A^{\frac{1}{3}} fm^{-2} $ then for a sufficiently
large A nucleus, $\Lambda_{QCD}^{2} \ll \mu^2$ such that
$\alpha_{s} ( \mu^2)\ll 1$  so that the method of weak coupling expansion
 would still be valid.In this scenario the soft gluon  fields produced
 by the color charges is classical and can be obtained by solving the 
 classical Y.M equations and from there they compute the distribution
 function given by,
 \be
 \frac{d~N}{d~ Y d^{2} k_{T}^2 \tau_{h}} =\frac {1}{\pi R_{A}^2} ( C_{F} N_{q} + C_{A} N_{g} )^{2}
\frac{2 g^{6} N_{c} }{(2 \pi)^{4}} \frac{1}{k^4} L( k_{T}, \lambda )
\ee
where
\be
L( k_{T}, \lambda ) \sim ln \left( \frac{k_{T}^2}{\lambda^2} \right) 
\ee
where  $\lambda$ is a cutoff scale associated with dynamical screening 
effects in dense partonic medium.  In the appropriate kinematical domain the predictions
of this model has been matched with that of pQCD results \cite{larry_nuclth}.
There has also been some effort to calculate the same using the 
coherent state description instead of the classical radiation formula, and they
were shown to agree with each other \cite{Matiniyan_Muller}
%%%%%%%%%%%%%%%%%%%%%%%%%%%%%%%%%%%%%%%%%%%%%%%%%%%%%%%%%%%%%%%%%%%%%%%%%%%%%%
\section{Space Time Evolution}
%%%%%%%%%%%%%%%%%%%%%%%%%%%%%%%%%%%%%%%%%%%%%%%%%%%%%%%%%%%%%%%%%%%%%%%%%%%%%%
\label{s_hyd_qrk}
 Once there are sufficient number of partons in the system, they will interact
 with each other and undergo a space time evolution.This space time evolution 
 of the QCD plasma, should be described by a gauge covariant transport
equations.Following (\cite{EGV}) one can define the gluon and quark
 distribution functions as.
\be
G_{\mu \nu }(x,p)= \int {d^{4}y\over (2\pi \hbar)^{4}}
e^{-\hbox{ip.y}/\hbar}
\left[\matrix{e^{-\frac{1}{2}\hbox{y.D}(x)}\,\vec{F}^{\lambda}_{\mu}(x)}\right]
\left[\matrix{e^{\frac{1}{2}\hbox{y.D}(x)}\,
\vec{F}_{\lambda \nu}(x)}\right]^{\dag},     
\label{gluwig}
\ee
\be
W(x,p)= \int {d^{4}y\over (2\pi \hbar)^{4}}
e^{-\hbox{ip.y}/\hbar}
\left[\matrix{e^{-\frac{1}{2}\hbox{y.D}(x)}\,\overline{{\Psi}(x)}}\right]
\left[\matrix{e^{\frac{1}{2}\hbox{y.D}(x)}\,{\Psi}(x)}\right],     
\label{qwig}
\ee
And use the equations of motion  to arrive at
the gauge covariant kinetic equations of Elze, Gyulassy and
Vasak.From there one can arrive at the classical non abelian transport equation,
after taking an ensemble average of the operator valued equation and then
setting terms of $O (\hbar =0 )$:,
\be
p^{\mu} D_{\mu} W \l( x,p \r) + g/2 p^{\mu} {\p_p}^{\nu} \l[
F_{\mu\nu},W \l(x,p \r) \r]_{+}=0
\label{trans1}
\ee 
Here $ [,]_+ $ means anticommutator. 
The distribution function W(x,p) ,
is a hermitian matrix in color space and for the SU(2) case can be written
in terms of a color singlet and triplet components as,
\be
W \l( x,p \r) = {\f {1}{2}} \langle G \rangle 1 +\f {1}{2} {\sum_{a=1}}^3
\lambda_a \lang G^a \rang 
\label{color_decomp}
\ee
Here $ \lang G \rang = Tr W \l( x,p \r)$  and $ \lang G_a \rang =
 Tr \l[ \lambda_a W \l( x,p \r) \r]$.  
Using Eq.~(\ref{color_decomp}) one can write Eq.~(\ref{trans1}) in terms of a
set of 
coupled partial differential equations.  The coupling is between
the singlet and triplet distribution function for quarks. They
are as follows, 
\be
\barr{rrr}
p_{\mu}\p^{\mu} \lang G \rang  +
g p_{\mu}{\p_{p}}^{\nu} \l[ {F_{\mu \nu}}^{a} \lang G^a \rang \r] & = & 0\\
p_{\mu}\p^{\mu} \lang G^{a} \rang + g \epsilon_{abc} p^{\mu}
{A^b}_{\mu} \lang G^c \rang +
g p_{\mu}{\p_{p}}^{\nu} \l[2 {F_{\mu \nu}}^{a} \lang G^a \rang \r] & = & 0\\
\earr
\label{colorkineticequation}
\ee
To take care of the  the effect of collisions  along with 
non perturbative production mechanism one needs to put a source  and a collision 
term on the right
hand side of this equation.   
\be
p^{\mu} D_{\mu} W \l( x,p \r) + g/2 p^{\mu} {\p_p}^{\nu} \l[
F_{\mu\nu},W \l(x,p \r) \r]_{+}= S_{ource} + C_{ollision}
\label{trans_soft}
\ee   
Lots of studies have been performed in the past ( see for instance references 
\cite{flux} to
\cite{Eisenber}) to investigate  the space time evolution of the plasma  
including  the effects of  source  and the collision term or just the 
source.The popular form of the source used in the  colorflux tube 
model is given as,
\begin{equation}
 S_{ource} = \sqrt{\left[E^{a}(\tau)E_{a}(\tau) \right]} \times ln \left[
1 -e^{ \left[- \pi \frac{ m^2 + p_{T}^2}{\sqrt{E_{a}(\tau) E^{a}(\tau) } } 
\right]} \right] 
\delta(p)
\end{equation}
And the collision term is usually written \cite{Eisenber} with the help of the 
equilibrium 
distribution
function $ W_{0}(p)$ and  the mean collision time $\tau_{0}$ as:,
\be
C_{ollision} = - \frac{W(x,p,t) - W_{0}(p)}{\tau_{0}}
\ee
\ni
In all these investigations it was observed that due to  
production of the particles the external field shows an exponentially damped 
oscillation
in proper time. From  the numerical solution of those equations 
one could see the feature of conversion of fields to particles and vice versa.
To take care of this time varying  external field, an artificial time dependence 
is introduced in the (Schwinger inspired) source term. 
Since in an exponentially decaying  external field, the tunneling picture of 
pair production is not appropriate, the  authors of \cite{ravi}  taking this 
into account derived a modified source term,using perturbation theory 
( with a decay constant of 0.1 fm). 
On the other hand if the produced particle current is strong, it would react back
on the system. To take care of the effect of quantum back  reaction by the 
particle current a  series of  investigations  were performed by 
F.cooper,E. Mottola et al. (\cite{cooper}).
And there it was pointed out, that the incorporation of the Pauli blocking 
in the  Schwinger inspired source term almost reproduces the exact field
 theoretical treatment. The pauli blocked source term they choose was of the form  
\begin{equation}
 S_{ource} = \left[ 1 - 2 W (x, p,t)\right] \sqrt{\left[E^{a}(\tau)E_{a}(\tau) 
\right]} \times ln \left[
1 -e^{ \left[- \pi \frac{ m^2 + p_{T}^2}{\sqrt{E_{a}(\tau) E^{a}(\tau) } } 
\right]} \right] 
\delta(p)
\label{pauli}
\end{equation}
The result of their \cite{cooper} exact calculation is shown in Figure 1. and 2.
where the solid line represents  the result of the  exact field theoretic 
calculation and the dotted line shows the result of incorporating 
 a Pauli-Blocked 
Source term (Eq. (\ref{pauli}) )to the Boltzman Vlasov equation.
\begin{figure}[ht]
\centerline{
\epsfig{figure=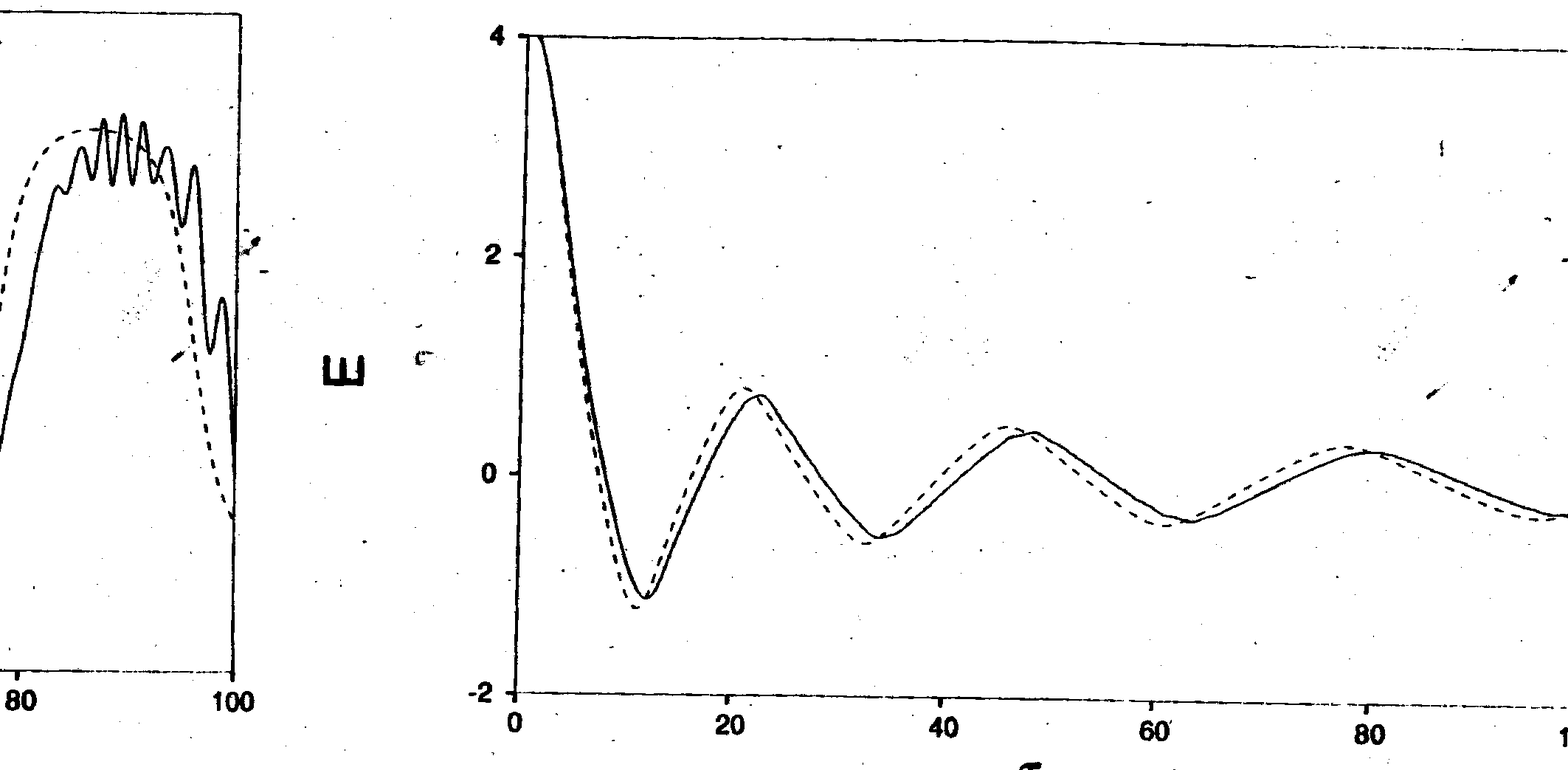,width=3cm}}
\caption{ Current vs time \hspace{79pt} Figure 2: Electric Field vs time }
\end{figure}
\noindent
%Going by their claim, it appears, that for a  more accurate phenomenological 
%study of the process, one can take the approach of \cite{Banerjee} Banerjee 
%et.al by just incorporating the  effect of Pauli blocking in their source term. 
 The half life of the field can be calculated from Fig. 2, 
and the value ( for a constituent quark mass) comes 
 close to that estimated in (\cite{ganguly3}), i.e $\sim$ 4 to 5 fm.
On the other hand for current quark mass, the same  number comes out to be
 240fm! After comparing these numbers it seems that the use of 
constituent quark mass is  more reasonable  for this model than the current
quark mass. 
\noindent 
The current in fig 1 is seen to  execute an oscillation in proper-time. The 
origin of this strong oscillation is the perhaps due to the fact that, in 
presence of an external field 
the negative and positively charged particles  tends to move away 
in opposite directions  hence creating a field with an opposite polarity.
 As the strength of this field, 
overcomes that of the external field these charges  moves in reverse 
direction and this prosess keeps  on continuing  generating  an oscillation in 
time.
% After few periods  of such  oscillations--because  of the collective 
%interaction--the plasma seems to 
%thermalise as is seen  in their plot of  entropy vs $\tau$.
The peak initial temperature attained, for constituent quark masses,in this
 model seems to be around $800~~MeV$, i.e well above the critical energy density required to 
produce the plasma.
%%%%%%%%%%%%%%%%%%%%%%%%%%%%%%%%%%%%%%%%%%%%%%%%%%%%%%%%%%%%%%%%%%%%%%%%%%%%%%
\subsection{Tunneling At Finite Temperature}
%%%%%%%%%%%%%%%%%%%%%%%%%%%%%%%%%%%%%%%%%%%%%%%%%%%%%%%%%%%%%%%%%%%%%%%%%%%%%%
In the previous section we have discussed how the field decays with time.
We also had noted from the study of Cooper et. al \cite{cooper}, that by the 
time the field has decayed to $\frac{1}{e}$th fraction of its original value,
the system has attained some temperature. Though this temperature is an 
oscillating function of proper time, still the time scale of oscillation is much
larger that the time scale of production. 
In ref.(\cite{ganguly3}),it was argued from the respective time scale analysis
of the relevant processes e.g production, depletion and thermalisation that, 
by the time the system thermalises, some residual mean field will still be 
present there in the system. The existence of such mean field for a fermionic 
as well as a bosonic system has also been noted in the studies of Blaizot and Iancu 
 \cite{Blaizot}. To see the effect of temperature, on Schwinger's tunneling 
mechanism,  an  investigation was carried out in \cite{ganguly3}, assuming the 
external filed to be  constant and homogeneous  every where; the effect of 
finite size  corrections was also taken into account in \cite{ganguly5}.For
 cyllindricaly symmetric system of radius R, the tunneling rate at a distance
$ \rho $ from the axis was found out to be:, 
\begin{eqnarray}
 &Im& F=\,\, \frac{1}{2}\displaystyle\sum_{n=1}^{\infty} 
 \left( \frac {gE}{n\pi} \right)^2
\sum_{p=1}^{\infty}
\left[ e^{\frac {-m^2~n\pi}{gE}-\frac {n\pi \left[\pi \left(
 p \right) + 3 \pi/4 \right]^2}{gER^2}}
% \,\, \times \right.
%\nonumber  \\
%&&
 %\left. 
 \left(\frac {2R}{
(p + 3/4)\pi^2 \rho} \times \right. \right. \nonumber \\
&&\left. \left. \,\,
Cos^2 \left[ \frac {\rho} {R} \left( p  +
\frac
{3}{4} \right) \pi - \frac {\pi}{4} \right]\right) \right]
-\frac{1}{2\pi}\sum^\infty_{n=1} (-1)^n \sum^\infty_{p=1}
 \frac {2R}{\pi^2 \rho
\left(
p +\frac {3}{4} \right)} \times \nonumber \\
&&\ Cos^2\left[\frac {\rho} {R} \left( 
 p + \frac {3}{4}  \right) \pi -  \frac {\pi} {4} \right]
 \displaystyle p.v \int^\infty_0 \frac {ds} {s^3} \left( (gES) Cot
(gES)  Sin \frac {n^2 \beta^2 gE} {4}\right) \nonumber \\
&& e^{-sm^2 - n^2\beta^2/4s}
 e^{-\frac {s \left[\pi p + \frac {3} {4} \pi \right]^2} {R^2}}
cosh \left( ng \beta \bar{A}_o \right) 
+ \frac {1} {2} \displaystyle\sum^\infty_{n=1} (-1)^n 
\sum^\infty_{p=1} 
\frac {2R}{\pi^2 \rho \left(
p +\frac {3}{4} \right)} \nonumber \\
&&\times Cos^2\left[\frac {\rho} {R} \left( 
 p + \frac {3}{4}  \right) \pi -  \frac {\pi} {4} \right]
 \displaystyle\sum^\infty_{\ell=1} \left( \frac {gE} {\ell \pi}
\right)^2
cosh \left( ng \beta \bar{A}_o \right)
%\nonumber \\
%&&
 Cos\left(\frac {n^2\beta^2 gE} {4}\right) \nonumber \\
 && e^{- \frac {\pi\ell m^2} {gE} - \frac {n^2\beta^2 gE} {4\pi\ell} -
\frac {\pi\ell \left[\pi ( p + 3/4)\right]^2} {gER^2}}
\end{eqnarray}\\
The Fig. 2a shows, how the rate varies with temperature and Fig.2b shows how
it varies as one moves radially outward.
% mass of the particles to $ 
%{m^2}_{eff}= \left( m^2 + \frac{ \left( \frac {3 \pi}{4} \right) } {R^2} \right)$      
\begin{figure}[h]
\centerline{
\epsfig{figure=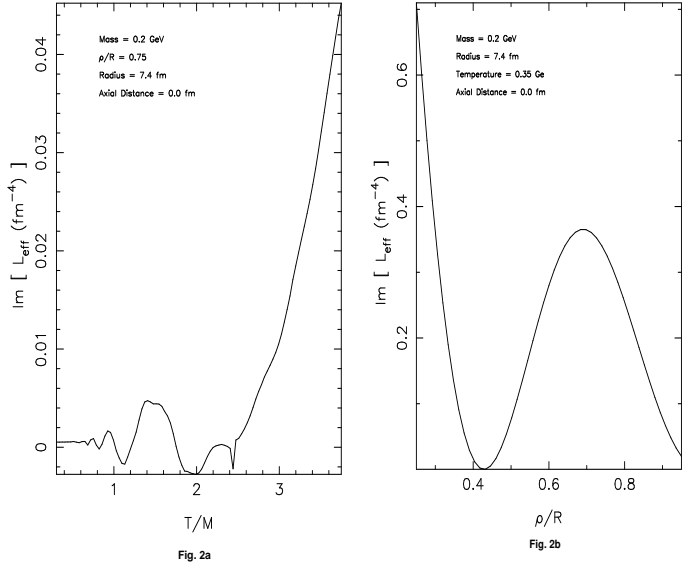,width=6cm}}
\caption{ (a) Production rate vs temperature. (b)Production rate vs axial distance}
\end{figure}
%Here R is the radius of the system and $\rho$ is the distance from the axis. 
%%%%%%%%%%%%%%%%%%%%%%%%%%% Delete %%%%%%%%%%%%%%%%%%%%%%%%%%%%%%%%%%%%%%%%%%%%
%We won't discuss it in any detail here, but the interested
%reader is requested to see \cite{geiger3}. We will discuss about the 
%predictions of this model later, and right now we'll take the opposite limit of 
%a collision dominated plasma to find out its predictions.\\
%%%%%%%%%%%%%%%%%%%%%%%%%%%%%%%%%%%%%%%%%%%%%%%%%%%%%%%%%%%%%%%%%%%%%%%
\subsection{Color Oscillation}   
In  the limit of very low number density and very weak external field, one can
study the collective oscillation of the plasma, by taking the momentum moments 
of equations (\ref{colorkineticequation})  and from there generating a set of 
color hydro-dynamic equations.
%  Below we will briefly describe procedure of how to generate a set
%of hydro-dynamical equations from the kinetic equations given by Eq( 
%\ref{colorkineticequation}). 
Before we go for generating the hydrodynamic equations, we normalize the 
4-velocities for quarks in the following way
\be
U^{\mu} U_{\mu} = 1 \\
~~~~~~~~{\hbox{ with }}~~~~~~\\
m U^{\mu} = \f {\int p^{\mu} G_a \l( x,p \r) d^4 p}{\int  G_a
\l( x,p \r) d^4 p} = \f {\int p^{\mu} G \l( x,p \r) d^4 p}{\int  G
\l( x,p \r) d^4 p}
\label{normalisation}
\ee
After taking the zeroth moment of eqn.(\ref{colorkineticequation}) we 
arrive at 
\be
  \p_{\mu} \l[ G^{\mu}\r] =0 
{\hspace{10pt} \hbox{and}\hspace{10pt}  }
  \p_{\mu}  \l[{G_a}^{\mu} \r] + g {\epsilon_{abc}} {A_\mu}^{b}
{G_c}^{\mu} =0 
\ee
Following Stewart \cite{stewart} if one decomposes the 4-vectors $ G^{\mu} =m n 
U^{\mu} $ and $ {G_a}^{\mu} =m {n_a}
 U^{\mu} $ one arrives at the mass and color continuity equations,
\be
\barr{ccc}
  \p_{\mu} \l[ {n} U^{\mu}\r] =0 
 {\hspace{10pt} \hbox{and}\hspace{10pt}  }
  \p_{\mu}  \l[U^{\mu} {n_a} \r] - g {\epsilon_{abc}} {A_\mu}^{b}
{n_c} U^{\mu} =0 &\\
\earr
\label{continuity1}
\ee
Further by defining the ratio of the two charge densities as $ Q^a = \f 
{n_a}{n}$ and using  eq.(\ref{continuity1}) one can obtain the color 
evolution equation , namely
\be 
  U^{\mu}\p_{\mu}  \l[{Q_a} \r] - g {\epsilon_{abc}} {A_\mu}^{b}
{Q_c} U^{\mu} =0 
\ee
Similarly from the first momentum moment of equation 
(\ref{colorkineticequation}) one arrives at,
\be
U^{\mu}\p_{\mu} U^{\nu} = {\f {g}{m}} {F_a}^{\mu\nu}U_{\mu}n_a = {\f 
{g}{m}}{F_a}^{\mu\nu}
{J_{\mu}}^{a}
\label{force-eqn}
\ee
Here repeated indices are summed up and ${J^{\mu}}_a = n U_{\mu}Q^a $ . 
{\footnote{ By
multiplying equation
(\ref{force-eqn}) by $U^{\nu}$ one can show 
 $U^{\mu}U_{\mu} =constant $;Normalising the U's appropriately 
Eqn.(\ref{normalisation}) is satisfied.}}    
These single flavour equations are  very similar  to the hydrodynamic equations
of Kajantie and Montonen \cite{kajantie} and they can easily be
generalised to the multiflavor 
case for quarks.   
\subsection{Gluon Hydro-dynamical Equations}
\ni To derive the gauge covariant  transport equations for Gluon one follows, 
the same procedure as before--starting from eqn( \ref{gluwig}).
%\be
%G^{ab}_{\mu \nu }(x,p)= \int {d^{4}y\over (2\pi \hbar)^{4}} 
%e^{-\hbox{ip.y}/\hbar}
%\left[\matrix{e^{-1/2\hbox{y.D}(x)} \, \vec{F}^{\lambda }_{\mu}(x)}
%\right]^{a}\left[\matrix{e^{1/2\hbox{y.D}(x)} \, 
%\vec{F}_{\lambda \nu}(x)}\right]^{b},
%\ee
\ni Upon using the equation of motion,and setting terms O($\hbar$) to zero
one arrives at the classical kinetic equation of the gluons,i.e,
\be
p^{\mu }.D_{\mu } G_{\mu \nu } + \frac{g}{2} P^{\sigma }\partial
^{\tau}_{p}\hbox{ $\left[ \right.$ }{\cal F}_{\sigma \tau},G_{\mu \nu
}\hbox{ $ \left. \right]_{+} $ }= g \pmatrix{{\cal F}_{\mu \sigma }G^{\sigma
}_{\nu }-&G_{\mu \sigma }{\cal F}^{\sigma }_{\nu }} \label{e_ke_glu}
\ee
Here $[,]_{+}$ means anti-commutator and
$D_{\mu } = \p_{\mu } - ig \left[\matrix{{\cal A}_{\mu },}\right]$ where, 
${\cal A}^{ab}_{\mu} \equiv - i f_{\hbox{abc}} A^{c}_{\mu}$, 
${\cal F}^{ab}_{\mu \nu} \equiv - i f_{\hbox{abc}} F^{c}_{\mu \nu} $and 
$f_{\hbox{abc}}$ is the antisymmetric structure constant for $SU(2)$ and $g$ 
is the coupling constant.
\ni In order to simplify things a bit further, one can assume  the quantity  
$G_{\mu \nu 
}$ to be symmetric w.r.t the Lorenz indices, which is technically known as spin 
equilibration ansatz, to write 
\be
G_{\mu \nu }(x,p) = p_{\mu } p_{\nu } G(x,p), \label{e_Gmn}
\ee
where $G(x,p)$ is a Lorentz scalar, but is a $3 \times 3$ matrix in color space. \
\ni One can check that with the ansatz of spin equilibration,  the right hand 
side of equation[\ref{e_ke_glu}]
vanishes and the Gluon kinetic equation in component (color) notation takes the 
form :
\bearr
p^{\mu }\partial_{\mu} G^{mn} + gp^{\mu }.A^{c}_{\mu}
\left[\matrix{f_{\hbox{cma}}G^{{an}}- G^{ma}f_{\hbox{can}}}\right]
+ i \frac{g}{2} p^{\sigma} \partial^{\tau}_{p} 
\hbox{ $\left[ \right.$}f_{\hbox{ema}}G^{{an}} +
f_{\hbox{ean}}G^{ma}\hbox{$ \left.\right] $ } 
\nonumber \\ \times F^{e}_{\sigma \tau} = 0 \hspace{10pt}
\label{gluonkinetic}
\eearr
In the above equation all repeated indices are summed over. 
From equation (\ref{e_Gmn}) we define the diagonal, antisymmetric and symmetric 
components as follows : 
\bearr
G^{11}  =  2 G^{1}, & 2 S^{1} = G^{23}+G^{32}, & 2i Q^{1} = G^{23}-G^{32} 
\nonumber \\
G^{22}  =  2 G^{2}, & 2 S^{2} = G^{31}+G^{13}, & 2i Q^{2} = G^{31}-G^{13} 
\nonumber \\
G^{33}  =  2 G^{3}, & 2 S^{1} = G^{12}+G^{21}, & 2i Q^{3} = G^{12}-G^{21} 
\eearr
\ni  We want to emphasize  here that the most  general the  hydrodynamic 
equations for Gluons  would not be the same as the quarks,it is an artifact
of the spin equilibration ansatz that they seem to come out to be the same here. 
In order to make some  more progress now we will further assume that all the 
symmetric combinations of the distribution function are zero. With this 
assumption one can show that the equations (\ref{gluonkinetic}) reduce to :
\bearr
p_{\mu}\p^{\mu}Q^{1}-gp^{\mu}\l[ A_{\mu}^2 Q_3 - A_{\mu}^3 Q_2 \r] +
g/2 p_{\mu}{\p_{p}}^{\nu} \l[ 2 {F_{\mu \nu}}^{1} \l( G^2+G^3 \r) \r] &
= & 0 \nonumber \\
p_{\mu}\p^{\mu}Q^{2}-gp^{\mu} \l[ A_{\mu}^3 Q_1 - A_{\mu}^1 Q_3 \r] +
g/2 p_{\mu}{\p_{p}}^{\nu} \l[ 2 {F_{\mu \nu}}^{2} \l( G^1+G^3 \r) \r] &
= & 0 \nonumber \\
p_{\mu}\p^{\mu}Q^{3}-gp^{\mu}\l[ A_{\mu}^1 Q_2 - A_{\mu}^2 Q_1 \r] +
g/2 p_{\mu}{\p_{p}}^{\nu} \l[ 2 {F_{\mu \nu}}^{3} \l( G^1+G^2 \r) \r] &
= & 0 
\label{gu1}
\eearr
and
\bearr
p_{\mu}\p^{\mu}G^{1} +
g/2 p_{\mu}{\p_{p}}^{\nu} \l[2 {F_{\mu \nu}}^{3} Q^3+{F_{\mu
\nu}}^{2} Q^2  \r] & = & 0 \nonumber \\
p_{\mu}\p^{\mu}G^{2} +
g/2 p_{\mu}{\p_{p}}^{\nu} \l[2 {F_{\mu \nu}}^{3} Q^3+{F_{\mu
\nu}}^{1} Q^1  \r] & = & 0 \nonumber \\
p_{\mu}\p^{\mu}G^{3} +
g/2 p_{\mu}{\p_{p}}^{\nu} \l[2 {F_{\mu \nu}}^{1} Q^1+{F_{\mu
\nu}}^{2} Q^2  \r] & = & 0 
\label{gu2}
\eearr
The set of equations for studying the hydrodynamic evolution of the plasma 
can be  obtained  as in the quark case after taking moments of the one particle
distribution function as before with the  only exception  that for the massless 
gluons one will have:,
\bearr
U^{\mu} U_{\mu} &=& 0.
\eearr
Instead of going into the  details of the derivation,
 we shall write down here final set of Gluon hydrodynamic equations:, 
The continuity equations are,
%On taking moments of eqns(\ref{gu1}) and (\ref{gu2}), and  defining 
%a quantity called
%color charge, as $ Q_a = \f {n_a}{n} $, that
\be
\p_{\mu} \non \l[ {n} U^{\mu}\r] =0 \hspace{20pt}
U^{\mu}\p_{\mu}  \l[{Q_a} \r] - g {\epsilon_{abc}} {A_\mu}^{b}
{Q_c} U^{\mu} =0
\label{continuity}
\ee
And the force balance equation is
\be
U^{\mu}\p_{\mu} U^{\nu} = {\f {g}{E}} {F_a}^{\mu\nu}
U_{\mu}Q^a
\label{forcebalance}
\ee
For studying the collective behavior of the system one needs to
solve Eq.( \ref{continuity}) and  Eq.( \ref{forcebalance}) along with the Yang 
Mills equation
\be
 D_{\mu}  {F_a}^{\mu\nu} = {J_a}^{\mu} = gnQ_a U^{\mu} 
\ee
self consistently. 
The basic idea here is that due to the self interaction the high momentum modes 
generate a low 
momentum long wavelength mean field which in turn acts back as a source term for 
a mean 
Yang-Mills field equation.In order to 
study the collective properties  one needs to solve these equations self
consistently, to describes the proper evolution of the system.
We have tried to solve these equations numerically, in 2+1 dimension. 
Initially susch a study was performed in \cite{bhat} in the  stationary frame of 
the plasma in 
(1+1) dimension, and it showed--depending upon the initial conditions---
a rich variety of regular as well as chaotic  solutions.
The motivation of going to one extra dimension was to understand how the system 
behaves under 
a small perturbation in one direction with an objective to understand the
role of chaotic dynamics in equilibrating the system. 
\begin{figure}[ht]
\centerline{
\epsfig{figure=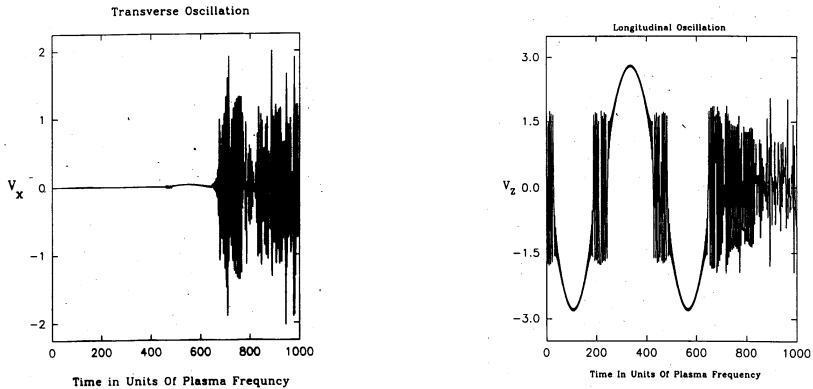,width=5cm}}
{\caption{Oscillation Profile:Left panel $V_{x} $ vs time.
 Right panel  $V_{z} $ vs time  } 
} \nonumber
 \end{figure}
\section{Study of Collective Oscillation of the Plasma}
\label{s_coll_osc}
The figures five and six shows the generic effect of a very small transverse
perturbation on the otherwise regular longitudinal oscillation.
As  can be seen from the figures(3a) and (3b) that upto time = $600 
\omega_p $ the 
velocity profiles remain same, but after that there is a catastrophic change
in  the velocity 
profile of $V_{x}$  and simultaneously the coherent oscillation in $V_{z}$ 
breaks up into a chaotic one.
A fast Fourier transform of these oscillations had shown that the most 
dominant 
frequency for both $V_x$ and $V_z$ components are the same, 
suggesting energy equilibration with the onset of chaotic motion. The most
intersting investigation here would be to see if this energy equilibration
is also accompanied by color equilibration.There have been some effort in
that direction by Gyulassy \cite{selikov}, but we will talk about it 
in the appropriate section.
%\nindent
Having described the role of soft dynamics in RHIC, we'll now briefly
introduce the corresponding picture in the still higher energy scale.
 That is the parton cascade model.
As one moves to higher and higher energy, it's believed that the hard processes
will dominate and produced partons will screen the long-range field. Therefore
to get the picture of partonic evolution in parton Cascade Model, one needs to
modify equation (\ref{trans_soft}), by putting the external field to be equal to
zero.Its worth noting that as a result of this, the two coupled partial
differential equations decouples from each other and one is left with:,
\be 
p^{\mu} \p_{\mu} W \,( x,p,t) = C_{ollision}  
\ee
This collision term describes, all the collision-al effects due to
$2 \rightarrow 2 $ and $2 \rightarrow 1$ and the reverse  processes, with 
appropriate modification of  distribution functions
due to pauli blocking and bose enhancement,
along with virtuality factors to take into account the space like and 
time like cascading.
The pathological low $p_t$ divergence is cured by  self consistently
introducing a screening mass for the partons.This selfconsistently modified 
parton cascade model is known as the self screened parton cascade model
\cite{SSPM} ,for the sake of brevity we  won't discuss it here, the details 
can be found in \cite{geiger3}. 
\subsection{Discussion and outlook}
Having introduced the  basic mechanisms and models of parton production in 
relativistic heavy ion collision, we will go over to the discussion of 
the prediction's of these models --valid at different energy scales.
As before we start from the low energy non-perturbative soft process
dominated regime. In the Boltzman Vlasov description with a Schwinger
like source term, it appears that the initial temperature at the onset of
equilibrium is dependent on the initial field strength that goes into the 
description. According to refrence \cite{bhalerao} for $ E_{0}^2 = 20 
GeV/fm^{3}$ the initial temperature at the onset of thermalisation
was found to be
varying between $ 225 $ to $ 250 $ MeV.
On the other hand according to reference ( \cite{cooper} fig. a constituent 
quark of mass .5 GeV the initial  peak temperature comes out to be around 4.5 
Gev, though for current quark mass, the temperature comes out much less. 
According to reference ( \cite{shu} ) the initial temperature in RHIC ( LHC) for 
gluons are estimated  to be 500 MeV ( 660 MeV) the same for quarks is 200 MeV ( 260 MeV) and the
thermalisation time for Gluons varies from 0.3 to 0.25 fm/c in RHIC and LHC 
energy scales. For gluons and light quarks the predictions from \cite{shu}
or \cite{jane} see to be close to each other and they suggest that the system 
undergoes a multistage thermalisation. The quarks are believed to be less 
equilibrated than gluons\cite{private}.
Although for  charm  or bottom quarks
the thermalisation time  $ \langle $ 2 fm/c.
%%%%%%%%%%%%%%%%%%%%%%%%%%%%%%%%%%%%%%%%%%%%%%%%%%%%%%%%%%%%%%%%%%%%%%%%%%%%%%%
The Initial Conditions as obtained from Self Screened Parton Model 
\cite{srivastav}
for Gold on Gold at BNL RHIC and  CERN LHC energies--prior to hydrodynamic 
expansion is given in Table I.
\begin{center}
Table I.
\end{center}
\begin{center}
\begin{tabular}{|c|c|c|c|c| }
\hline Self Screened Parton Cascade &  $\tau_{i} $(fm/c)  & $T_{i}$ & 
$\epsilon_i$ $( GeV /fm^3 )$  \\
\hline RHIC& 0.25 &0.668 & 61.4  \\
\hline  LHC& 0.25 & 1.02 & 425   \\
%\hline 4&xxx &-&-&1&-  2 ML &\\
\hline
\end{tabular}
\end{center}
\noindent
Coming back to the question of color equilibration, though according to 
\cite{selikov}, color equilibrates takes place before the system reaches thermal 
equilibration, but  it would be nice to demonstrate this result using the color
kinetic equations, retaining their full nonlinearity. It is generally felt that
such an investigation would shed light on the question, whether the onset of 
chaotic oscillation has really got some thing to do with the equilibration 
procedure or not. And lastly it's worth mentioning that 
all the previous studies reveal that in RHIC the system would not 
come chemical equilibrium though at LHC energy scale it might get very close to
it. 
\section{Acknowledgement}
The author would like to thank Prof. B.C Sinha for  providing him with the local 
hospitality of SINP Calcutta--during which a part of this work was done. He also
would like to thank Dinesh  K. Srivastava and Rajiv Bhalerao  for numerous 
illuminating communications on various issues related to this  work. Lastly
I would like to thank S.Lal for his invaluable help while preparing the
manuscript.


\begin{thebibliography}{[99]}
\bibitem{satz}     H. Satz et.al., eds., Proc. Sixth Intern. Conf. on Ultra 
relativistic nucleus-nucleus collisions, Quark matter 1987 ( Nordkirchen, August
1987), Z.Phys. C38 (1988) 1.
\bibitem{Low}   F.E. Low, Phys. Rev.D12 (1975) 163; S. Nussinov Phys. Rev. Lett. 
34 (1975) 1286
\bibitem{ganguly1}   A.K. Ganguly Ph. D Thesis submitted to M.S.University 
(1993);unpublished and the references there in. 
\bibitem{Schwinger}  J. Schwinger, Phys. Rev.~82.(1951) 664
\bibitem{Gyu}           M. Gyulassy and C.Iwazaki Phys .Lett {\underline{B165}},    
                     (1985) 157;
\bibitem{finitesize}   C. Martin And  D. Vautherin Phys. Rev. {D38}, (1988)3593 
; Phys. Rev. \underline{D40}, 1667 (1989);C.S. Warke and R.S. Bhalerao, Pramana 
- J. Phys. 
\underline{38}, 37 (1992).
\bibitem{ganguly2}     A.K.Ganguly, P.K. Kaw  And J. C. Parikh, Phys. Rev        
                      {\underline{D48}} (1993) R2983
\bibitem{bhalerao}     B. Banerjee, R.S. Bhalerao and D. Ravishankar, 
Phys.Lett.\underline{B224}, 16 (1989)

\bibitem{eskola}      K.J. Eskola, hep-ph/9708472 and references therein.
\bibitem{saturation}         J.P. Blazot and A. Muller, Nucl. Phys. \underline{B 
                             289}(1987) 847; L.V. Gribov, E.M. Levin and M.G.    
                             Ryskin, Phys.Rep.100 (1983) 1; K.J  Eskola and      
                             Kajantie, Z. Phys. {\underline{C75}},(1997),515.

\bibitem{larry}     L. Mclerran and R. Venugopalan, Phys. Rev.{\underline{D49}}  
                     (1994) , 2233; \& 3352; {\em{ibid}}, {\underline{D50}} (1994),      
                     2225.ibid,{\underline{53}} (1996), 458.

\bibitem{larry_nuclth}   M. Gyulassy and L. McLerran, nucl-th/9704034; Phys. Rev
                      {\underline{C~56}}, (1997), 2219. J.F Gunion and G.        
                      Bertsch, Phys. Rev. {\underline{D~25}}, (1982), 746

\bibitem{Matiniyan_Muller} Y.V. Kovchegov and D.H. Rischke Phys.Rev.
{\underline{C~56}}, (1997),1084. S. Matinyan B. Muller \& D.H. Rischke, Phys. 
Rev.{\underline{C~56}}, (1997), 2191.


\bibitem{EGV} H. T. Elze, M. Gyulassy and D. Vasak, Nucl. Phys.
            \underline{B276}, (1986), 706
            H. T. Elze, M. Gyulassy and D. Vasak,  Phys. Lett. 
            \underline{ 177B}, (1986), 402;
             D. Vasak M. Gyulassy and H. T. Elze,  Ann.  Phys.(N.Y) ,  
            \underline{ 173}, (1987), 462; H. -Th. Elze and U. Heinz, Phys.Rep
            183, (1989),81;U. Heinz,  Ann. Phys.(N.Y), 161, (1985), 48;H.T. Elze 
and U. Heinz, in {\it Quark Gluon Plasma}, Ed. R.Hwa
             World Scientific, Singapore, 1990)
            
\bibitem{flux}Andersson et.al,  Phys.Rept. 97 (1983) 31;Biro et.al, Nucl. Phys. 
{\underline{B245}} (1984)449;A.K.~Kerman T. Matsui and B. Svetitsky, Phys.Rev.Lett. 56 (1986) 
219;Kajantie and Matsui ,Phys.Lett.{{\underline164B}} (1985) 373;
A. Bialas and W. Czyz, Phys. 
Rev. \underline{D31}, 198 (1985);G. Gatoff, A.K. Kerman and T. Matsui, Phys. 
Rev. {\underline{D36}} (1987) 114; N.~K. Glendenning and T. Matsui, $ibid.$ 28,
(1983), 2890 B. Banerjee, R.S. Bhalerao and D. Ravishankar, Phys. 
Lett.\underline{B224}, 16 (1989); Bialynicki-Birula et.al., Phys.Rev. 
\underline{D44} (1991) 1825;

\bibitem{taozero}       Kajantie \& Matsui, Phys. Lett. \underline{B164} (1985) 
373;                      Asakawa \& Matsui, Phys. Rev. \underline{D43} (1991) 
2871.

\bibitem{Eisenber}    Eisenberg, hep-ph/9609205, Found. Phys. 27 (1997) 1213.
                                     
\bibitem{ravi}        Ravishankar and Bhalerao, Phys. Lett,{\underline{409}}, 
(1997) 38 .               
\bibitem{cooper}      Cooper et al. Phys. Rev. \underline{D48} (1993) 190; Y. Kluger et. al
Phys. Rev. Lett. 67,(1991) 2427. For a recent discussion one can look at
Y. Kluger, E. Mottola and J. M. Eisenberg, hep-th/9803372.

\bibitem{Blaizot}          J.P. Blaizot and E. Iancu, Nucl. Phys \underline{B417},   
608(1994);Phys. Rev. Lett. 72,3317,(1994) 

\bibitem{ganguly3}   A.K. ganguly, P.K. Kaw and J.C. Parikh, Phys. Rev.
{\underline{C51}} (1995) 2091. 
\bibitem{ganguly5}  A.K Ganguly  in {\it  Physics  and Astrophysics Of Quark 
-Gluon Plasma}, 
Ed. Bikash C. Sinha et al. 
             Narosa Publishing House,New Delhi,1997),

\bibitem{geiger3}    K.Geiger \& B.Muller, Nucl. Phys. {\underline{B369}}(1992), 
600; K.Geiger, Phys. Rev. {\underline{D46}}, (1992),4965; 
{\underline{D46}}, (1992), 4986;Phys. Rep. 258 (1995),237
\bibitem{stewart} J. M. Stewart, in {\it Non-Equilibrium Relativistic
Kinetic Theory}, Springer-Verlag, Berlin/New York, 1971.
\bibitem{kajantie} K. Kajantie and C. Montonen, Physica Scripta, \underline{
22}, (1981), 555  

\bibitem{bhat} J. Bhat, P.K. Kaw and J.C. Parikh, Phys. Rev. \underline{
                D39}, (1989), 646

\bibitem{selikov} A. V. Selikhov and M. Gyulassy Phys. Lett. \underline{B316}, 
(1993)373. 
\bibitem{SSPM}     Muller, Biro, et. al on SSPM.
\bibitem{G} A.K. Ganguly, P. K. Kaw and J. C. Parikh,(1989); unpublished 
\bibitem{shu}  E. Shuryak, Phys. Rev. Lett. 68, (1992), 3270.
\bibitem{jane}  J.Alam, S. Raha, B.Sinha Phys. Rev. Lett. 73,
(1994), 1895.
\bibitem{private}    D.K. Srivastava , Private communication.
\bibitem{srivastav}  D.K. Srivastava Munshi Golam Mustafa and B. Muller,
                   Phys. Rev {\underline{C56}},(1997),1064.
\end{thebibliography}
\end{document}